\documentclass[12pt,preprint]{aastex}

\def\gtsima{$\; \buildrel > \over \sim \;$}
\def\simgt{\lower.5ex\hbox{\gtsima}}

\shorttitle{GOODS~850-5 lies most probably at z$\sim$4}
\shortauthors{Dannerbauer et al.}

\begin{document}


\title{Interferometric detections of GOODS~850-5 at 1~mm and 1.4~GHz\altaffilmark{1}}


\author{H. Dannerbauer, F. Walter}
\affil{Max-Planck-Institut f\"ur Astronomie,
              K\"onigstuhl 17, D-69117 Heidelberg, Germany; email: dannerb@mpia-hd.mpg.de,walter@mpia-hd.mpg.de}

\and
\author{G. Morrison}
\affil{Institute for Astronomy, University of Hawaii, Honolulu, HI 96822, USA; email: morrison@ifa.hawaii.edu}
\affil{Canada-France-Hawaii Telescope, Kamuela, HI, 96743, USA;
email: morrison@cfht.hawaii.edu}




\altaffiltext{1}{Based on observations carried out with the IRAM Plateau de
Bure Interferometer. IRAM is supported by INSU/CNRS (France), MPG (Germany)
and IGN (Spain). The National Radio Astronomy Observatory is a facility of the
National Science Foundation operated under cooperative agreement by Associated
Universities, Inc.}

\begin{abstract}
We have obtained a position (at sub-arcsecond accuracy) of the submillimeter
bright source GOODS~850-5 (also known as GN10) in the GOODS North field using
the IRAM Plateau de Bure interferometer at 1.25~mm wavelengths
(MM~J123633$+$6214.1, flux density: $S_{1.25~mm}=5.0\pm1.0$~mJy). This source
has no optical counterpart in deep ACS imaging down to a limiting magnitude of
i$_{775}=28.4$~mag and its position is coincident with the position found in
recent sub-millimeter mapping obtained at the SMA \citep{wan07}. Using deep
VLA imaging at 20~cm, we find a radio source ($S_{20~cm}=32.7\pm4.3~\mu$Jy) at
the same position that is significantly brighter than reported in Wang et
al. The source is detected by Spitzer in IRAC as well as at
24~$\mu$m. We apply different photometric redshift
estimators using measurements of the dusty, mid/far-infrared part of the SED
and derive a redshift z~$\sim4$. Given our detection in the millimeter and
radio we consider a significantly higher redshift \citep[e.g.,
z~$\sim6$][]{wan07} unlikely. MM~J123633$+$6214.1 alias GOODS~850-5
nevertheless constitutes a bright representative of the high-redshift tail of
the submillimeter galaxy population that may contribute a significant fraction
to the (sub)millimeter background.
\end{abstract}


\section{Introduction}
Research in the last decade showed that to understand the assembly and
formation of massive galaxies it is critical to study Submillimeter Galaxies
\citep[SMGs; see][for a review]{bla02}. Since their first detection ten years
ago \citep{sma97}, more than three hundred dust-enshrouded high-z objects (z
$>$2) have been detected by SCUBA and by MAMBO in the submillimeter and
millimeter regime
\citep[e.g.,][]{hug98,ber00,dan02,dan04,sma02,gre04,wan04,pop06,cop06,ber07}. Due
to the large beam size in the (sub)mm (SCUBA: 15\arcsec, MAMBO: 11\arcsec), it
is impossible to do a proper identification in other wavebands (such as the
optical and near-infrared) based on bolometer data only. Thus, typically
interferometric observations at radio wavelength with the VLA were used to
identify the location of the SMGs, yielding a radio identification rate of
$\sim50-70$\% \citep[e.g.,][]{sma00,dan04,wan04,pop06,ivi07}, depending on
both the depths of the radio and the bolometer maps. The main drawback of this
technique is, however, that the radio flux is redshift dependent, whereas the
(sub)mm flux density at z~$\ge1$ is not \citep[due to the inverse K
correction; e.g.,][]{bla93}: one would thus expect the highest redshift
(z~$>3$) sources \citep[e.g.,][]{bar00} to be very faint at radio wavelengths
if they were to follow the radio-FIR correlation \citep[e.g.,][]{con92}. We
note that the radio-identified SMG with the highest redshift lies at
z~$\sim3.6$ \citep{cha03,cha05}. Based on this sample, CO observations of
radio identified SMGs \citep[e.g.,][]{ner03,gen03,gre05} lead to the main
conclusion that SMGs are massive galaxies in formation and \citet{cha05} argue
that the contribution to the star-formation rate density (SFRD) from this
population is missed by UV-selected surveys.  If one only uses
radio-identified sources \citep[e.g.,][]{cha03,cha05} one thus has to be aware
of the fact that the radio selection introduces a bias towards the low end of
the true redshift distribution. To summarize, our knowledge about the nature
and properties of the SMG population is mainly based on radio-identified SMGs
--- a biased sample.

One of the open questions in the investigations of SMGs is if there exists a
significant very high-redshift tail \citep[z~$>4$;
e.g.,][]{dan02,dan04,val07,you07}.  The existence of such a tail would not be
in contradiction to constraints from measurements of the cosmic infrared
background \citep{gis00}. Interferometric observations in the (sub)mm hold
the promise to locate the positions of bright SMGs to sub-arcsec
accuracy. One of the brightest SCUBA sources in the GOODS North field is
GOODS~850-5 (Wang et al. 2004; this source is labeled GN~10 in Pope et
al. 2005\footnote{The distance between these two positions is 2.8\arcsec\/,
i.e. within the SCUBA bolometer position uncertainties.}) with a submm flux
$S_{850~\mu m}=12.94\pm2.14$~mJy \citep{wan04}\footnote{\citet{pop05} report
$S_{850~\mu m}=11.3\pm1.6$~mJy}. Until very recently this bright SMG did not
have a known counterpart which motivated observations with the IRAM Plateau de
Bure Interferometer \citep[PdBI;][]{gui92} to obtain an
accurate position for this source. In the meanwhile, \citet{wan07} reported
the location of GOODS~850-5 through SMA observations. Based on their detection
in the submm (and subsequent identification in IR observations) and a
2~$\sigma$ detection in the radio ($S_{1.4~GHz}=18.7\pm8.6~\mu$Jy),
\citet{wan07} argue that the redshift of GOODS~850-5 is z~$>4$ and could even
lie between z~$=5.6-8.0$ - with a much lower probability at z$<4.5$.

Here we report the results based on our PdBI observations.  We adopt the
cosmological parameters $\Omega_{matter}=0.27, \Omega_{\Lambda}=0.73,$ and
$H_{0}=71~km~s^{-1}~Mpc^{-1}$ \citep{spe03,spe07}. Furthermore, we assume a
Salpeter initial mass function from 0.1 to 100~$M_{\sun}$. We use the AB
magnitude scale \citep{oke83}.

\section{Observations}
The final map resulting from our observations between December 2005 and August
2006 with the Plateau de Bure millimeter interferometer  in C and
D configuration are based on  6.8~hours of on-source observing time. The
phase center (see Table \ref{tab:position}) was set at the nominal SCUBA
position published in \citet{wan04}. The source was observed in the 1~mm and
3~mm band simultaneously. The 1~mm receiver was tuned at 240.000~GHz
(corresponding to 1.25~mm) to obtain an accurate position. The 3~mm receiver was
tuned to 98.3188~GHz, i.e., the frequency of the CO(2-1) transition redshifted to z~$=1.3448$.
This is the redshift of the optical source GOODS850-5a \citep{wan04} which is
situated close ($\sim2.4\arcsec$) to the SCUBA bolometer position. The total
bandwith of 580~MHz covers a redshift space of $\Delta z=0.0138$ at that
redshift (i.e. within the redshift uncertainty of the source, cf: z~$=1.344$
in Cohen et al. 2000 and Pope et al. 2006; z~$=1.345$ in Wang et al. 2004;
z~$=1.34476$ in Wirth et al. 2004).

The data were calibrated through observations of standard bandpass (3C273,
J0418$+$380), phase/amplitude (1044$+$719, 1150$+$497) and flux calibrators
(CRL~618, 3C273, 1044$+$719, 1150$+$497) and reduced with the GILDAS software
packages CLIC and MAP. The FWHM of the beam is $1.4\arcsec\times1.1\arcsec$ at
240~GHz and $3.4\arcsec\times3.2\arcsec$ at 98.3~GHz.

We detected the dust continuum emission of GOODS~850-5 at 5 $\sigma$
significance at $S_{1.25~mm}=5.0 \pm 1.0$~mJy. The derived position is
RA(J2000)$=$12:36:33.45, DEC(J2000)$=$\- 62:14:08.95 (hereafter:
MM~J123633$+$6214.1) with an estimated position uncertainty of
$\pm$0.3\arcsec\/. The position and flux density of MM~J123633$+$6214.1 are
derived from point source fits of the visibilities in the UV plane and are
consistent with the values derived in the image plane. The difference between
the SCUBA \citep{wan04} and our millimeter interferometric position is only
--0.02~\arcsec\/ in RA and --0.48~\arcsec\/ in DEC, and our position is
consistent with the SMA position from \citet{wan07}, see summary in
Table~\ref{tab:position}. In the 3~mm band no line was detected down to a rms
of 0.8~mJy~beam$^{-1}$ (channel width: 40~MHz, 120~km~s$^{-1}$) on the
position of the optical source at z~$=1.3448$ (and also at the position of
GOODS~850-5). This corresponds to an upper limit (3~$\sigma$) of the
integrated flux intensity of 1.2~Jy~km~s$^{-1}$, adopting a typical CO width
for a SMG of 500~km~s$^{-1}$ \citep[e.g.,][]{gre05}. However, we note that a
number of CO detections of SMGs have even lower integrated flux intensities
than this 3~$\sigma$ upper limit \citep[e.g.,][]{gre05}.

We have used new, deep VLA radio continuum observations (Morrison et al. 2008)
to constrain the radio flux density of GOODS~850-5. These observations reveal
a highly significant radio source ($S_{1.4~GHz}=32.7\pm4.3~\mu$Jy); i.e. a
source that is significantly brighter than the upper limit reported in
\citet{wan07}. Our derived flux density is however in agreement with the
3~$\sigma$ detection reported in \citet{pop06}. The radio emission is
unresolved with a FWHM~$\sim1.7\arcsec$.

In our analysis, we use the exquisite multi-wavelength dataset available for
the GOODS North field (e.g.,
Alexander et al. 2003; Giavalisco et al. 2004; Dickinson et al., in
preparation) and have corrected these data products for the offset of
--0.38\arcsec\/ in declination between VLA and the GOODS imaging products
\citep[see also][]{pop06}.

\section{Discussion}
\subsection{Multi-wavelength Properties of MM~J123633$+$6214.1 alias GOODS~850-5}
The PdBI/VLA source is undetected in the deep ACS images down to
$i_{775}=28.4$~mag. However GOODS~850-5 is detected at low significance in
Spitzer IRAC measurements at the few $\mu$Jy level \citep{pop06} as summarized
in Table \ref{tab:flux}. The source is also detected at 24~$\mu$m
(Table~\ref{tab:flux}) but undetected both at MIPS 70~$\mu$m and 160~$\mu$m at
a 3~$\sigma$ level ($S_{70~\mu m}<2.0$~mJy and $S_{160~\mu
m}<15.0$~mJy). However, due to the large beam size (18\arcsec\/ at 70~$\mu$m and
40\arcsec\/ at 160~$\mu$m) our source would be confused with other objects
within the beam. We conclude that the IRAC counterpart presented in
\citet{pop06} is the proper counterpart of MM~J123633$+$6214.1\footnote{The
IRAC and MIPS 24~$\mu$m fluxes by \citet{pop06} are consistent within
1.5~$\sigma$ with the measurements reported by \citet{wan07}}.

Due to the lack of a spectroscopic identification, we apply several
photometric redshift estimators focusing on different parts of the SED in
order to put constraints on the possible redshift--range of the
source. Applying the radio/(sub)millimeter index \citep{car99,car00} on the
PdBI flux and the radio flux we estimate a redshift
z$=3.81^{+2.04}_{-1.25}$. Furthermore, we use the infrared galaxy SED
templates from \citet{char01} to fit the SED of our source (using the
MIPS~24~$\mu$m, SCUBA, PdBI and VLA flux) and this method proposes a redshift
of z~$\sim3.3\pm1.0$, --- within the errors consistent with the radio/(sub)mm
spectral index estimates --- and infrared luminosity
$L_{IR}\sim1\times10^{13}L_{\sun}$. We obtain the same result by only using
the MIPS~24~$\mu$m and the VLA flux. The use of the 850/1.2~mm ratio as a
proper redshift indicator is very limited \citep[see Fig.~13
in][]{gre04}. However, the SCUBA to PdBI flux ratio of 2.6 limits the possible
redshift range from z~$\sim0-4$ \citep[see plot in][]{gre04}. As \citet{wan07}
we have also applied the widely used photometric redshift code {\it hyperz}
\citep{bol00} on the optical/near-infrared data from ACS and IRAC. Our
redshift estimates give a redshift z~$\sim6$, i.e. consistent with the range
of values proposed by \citet{wan07}. However, we note that with this method
only the stellar component of the source is fitted and the applied templates
do not reflect the characteristic properties of infrared galaxies.

Assuming a redshift z~$\sim4$, taking into account the optical/near-IR
measurements from ACS/Spitzer and applying the reddening law by \citet{cal00}
we estimate the extinction to be A$_{V}\sim3$. For comparison, \citet{sma04}
and \citet{swi04} derived for SMG samples mean reddening of A$_{V}\sim1.70$
and 3.0, based on optical/near-IR colors.

The source is undetected both in the soft (0.5-2~keV) and hard (2-8~keV) bands
in the 2~Ms Chandra observations of GOODS North
\citep[e.g.,][]{ale03}. However, the lack of X-ray detection cannot exclude
the presence of an heavily obscured AGN in GOODS~850-5. Using IRAC,
\citet{ivi04} proposed a diagnostic tool in order to distinguish between AGN
and starburst as dominating energy source. Applying this tool on
MM~J123633$+$6214.1 would propose an AGN at redshift z~$\sim2$. However, we
note that \citet{lac04} and \citet{ste05} pointed out that starbursts at
z~$\sim4$ contaminate the IRAC selection for z~$\sim1-2$ AGNs.

Using either the template fitting from \citet{char01} (resulting in an
infrared luminosity $L_{IR}\sim1\times10^{13}L_{\sun}$ and taking into account
the conversion between infrared luminosity $L_{IR}$ and SFR \citep{ken98}), we
derive a star formation rate SFR~$\approx 1800~M_{\sun}~yr^{-1}$. By simply using
the PdBI flux of 5.0~mJy at 1.25~mm we obtain a star formation rate
SFR~$\approx 2200~M_{\sun}~yr^{-1}$ \citep{ber03}. This is consistent with the
SFR derived from the radio flux for a source at z~$\sim4$ of SFR~$\approx
2000~M_{\sun}~yr^{-1}$ (see discussion in Ivison et al. 2002).

We conclude that the multi-wavelength data suggests that MM~J123633$+$6314 is
a starburst at z~$\sim$4.

\subsection{Radio Faint SMGs in GOODS North}
A total of 21 SMGs in GOODS North are securely identified \citep{pop06}. In
addition to MM~J123633$+$6214.1 there are four SMGs that have even fainter VLA
20~cm fluxes ($\leq 33~\mu$Jy) than our source. Similar to the counterpart of
MM~J123633$+$6214.1 none of them has an ACS counterpart and their IRAC fluxes
are faint ($\leq 20~\mu$Jy). One of these SMGs is the well-studied source
HDF850.1 \citep{hug98,dow99,dun04} which is one of the excellent candidates to
lie at z~$\sim4$ or higher. To summarize, about 20\% of securely identified
SMGs could lie at high redshift z~$\sim4$ or higher. \citet{dan02,dan04} and
\citet{you07} also reported the discovery of reliable candidates of redshift
z~$\sim4$ or higher in other deep fields.
  
\section{Conclusion}
We have used the IRAM Plateau de Bure Interferometer to obtain an accurate
position and flux at 1.3~mm wavelengths of the SMG MM~J123633$+$6214.1 alias
GOODS 850-5 ($S_{1.25~mm}=5.0\pm1.0$~mJy). This source is coincident with a
source detected in deep VLA radio continuum observations at high significance
($S_{1.4~GHz}=32.7\pm4.3~\mu$Jy). The source has also been detected by recent
observations obtained at the SMA \citep{wan07} and by Spitzer \citep{pop06} in
the IRAC and 24~$\mu$m bands. The photometric redshift indicators using the
MIR- and FIR SED of MM~J123633$+$6214.1 all point towards a redshift of
z~$\sim4$. Based on our millimeter and radio detection, we consider a much
higher redshift (z~$\sim6$, as proposed by Wang et al. 2007)
unlikely. However, only future facilities with wide mm-receivers planned
e.g. the upgraded PdBI or the LMT will unambiguously solve the question of the
true redshift of GOODS~850-5 (ALMA will unfortunately not be able to reach
GOODS North).  In any case, GOODS~850-5 represent one of the few solid
candidates of a high-z (z~$>3$) SMG that contribute to the cosmic SFRD at
early cosmic times.

\acknowledgments
We would like to thank the IRAM staff for help provided during the
observations and especially Arancha Castro-Carrizo and Roberto Neri for
helpful support during the data reduction at IRAM Grenoble. Furthermore, we
are grateful to Dave Frayer for 70 and 160~$\mu$m measurements on the PdBI
position and Emanuele Daddi for instructive conversations and help in applying
the \citet{char01} SED templates on GOODS~850-5. Finally, we appreciate the
comments of the anonymous referee which clarified a number of points and
improved our manuscript.

{\it Facilities:} \facility{IRAM:Interferometer}, \facility{VLA}, \facility{HST (ACS)},
\facility{Spitzer (IRAC, MIPS)}, \facility{CXO (ACIS)}.






\clearpage
\begin{table*}
\tiny
\caption{Positions of the SMG MM~J123633$+$6214.1 alias GOODS~850-5\label{tab:position}}
\begin{tabular}{lcllcl}
\hline\hline
ID&Instrument&R.A.&Decl.&Offset&Reference\\
&&(J2000.0)&(J2000.0)&$\Delta PdBI-Other$&\\
\hline
MM~J123633$+$6214.1&PdBI&12:36:33.45$\pm$0.04&62:14:08.95$\pm$0.3&---&{\bf this paper}\\
\hline
VLA Counterpart & VLA & 12:36:33.42$\pm$0.01& 62:14:08.70$\pm$0.1&0.33\arcsec&{\bf this paper}\\
GOODS~850-5&SMA& 12:36:33.45$\pm0.03$&62:14:08.65$\pm0.2$&0.30\arcsec&\citet{wan07}\\
GOODS~850-5&SCUBA&12:36:33.45$\pm$0.73&62:14:09.43$\pm$5.1&0.48\arcsec&\citet{wan04}\\
GN~10
(SMMJ123633.8$+$621408)&SCUBA&12:36:33.8$\pm$1.14\tablenotemark{\Diamond}&62:14:08$\pm$8.0\tablenotemark{\Diamond}&2.63\arcsec&\citet{pop05}\\
\hline
\end{tabular}
\tablecomments{
Units of right ascension are hours, minutes, and seconds, and units of
declination are degrees, arcminutes, and arcseconds.}
\tablenotetext{\Diamond}{The given
position uncertainty is the search radius for counterparts of 8\arcsec\/ applied by \citet{pop06}.}
\end{table*}

\clearpage

\begin{table*}
\tiny
\caption{Fluxes of the SMG MM~J123633$+$6214.1 alias GOODS~850-5\label{tab:flux}}
\begin{tabular}{lrccl}
\tableline\tableline
Band&Unit&MM~J123633$+$6214.1&Instruments&Reference\\
(1)&(2)&(3)&(4)&(5)\\
\tableline
$X-Ray_{0.5-2~keV}$&$10^{-17}$~ergs~cm$^{-2}$~s$^{-1}$&$<$2.5&Chandra&\citet{ale03}\\
$X-Ray_{2-8~keV}$&$10^{-16}$~ergs~cm$^{-2}$~s$^{-1}$&$<$1.4&Chandra&\citet{ale03}\\
$B_{435}$&mag&$>$29.1&ACS&\citet{gia04}\\
$V_{606}$&mag&$>$29.1&ACS&\citet{gia04}\\
$i_{775}$&mag&$>$28.4&ACS&\citet{pop06}\\
$z_{850}$&mag&$>$27.9&ACS&\citet{gia04}\\
$S_{3.6~\mu m}$&$\mu$Jy&1.21$\pm$0.39&IRAC&\citet{pop06}\\
$S_{4.5~\mu m}$&$\mu$Jy&1.96$\pm$0.36&IRAC&\citet{pop06}\\
$S_{5.8~\mu m}$&$\mu$Jy&2.72$\pm$0.88&IRAC&\citet{pop06}\\
$S_{8.0~\mu m}$&$\mu$Jy&5.11$\pm$1.14&IRAC&\citet{pop06}\\
$S_{24.0~\mu m}$&$\mu$Jy&30.7$\pm$5.4&MIPS&\citet{pop06}\\
$S_{70~\mu m}$&mJy&$<$2.0&MIPS&\citet{huy07,fra07}\\
$S_{160~\mu m}$&mJy&$<$15.0&MIPS&\citet{huy07,fra07}\\
$S_{850~\mu m-Wang04}$&mJy&12.94$\pm$2.14&SCUBA&\citet{wan04}\\
$S_{850~\mu m-Pope06}$&mJy&11.3$\pm$1.6&SCUBA&\citet{pop06}\\
$S_{870~\mu m}$&mJy&12.0$\pm$1.4&SMA&\citet{wan07}\\
$S_{1.25~mm}$&mJy&5.0$\pm$1.0&PdBI&{\bf this paper}\\
$S_{1.4~GHz}$&$\mu$Jy&32.7$\pm$4.3&VLA&{\bf this paper}\\
\hline
\end{tabular}
\tablecomments{
Col. (1): Band in which flux is measured. Col. (2): Units of the flux
density measurements. Limits are 3~$\sigma$. Optical/near-IR magnitudes are
measured in a 0.2\arcsec\/ diameter aperture and are on the AB system
\citep{oke83}, see also \citet{gia04}. IRAC fluxes are measured in a
4\arcsec\/ diameter aperture. MIPS 24~$\mu$m flux is measured within a
5.7\arcsec\/ aperture. For the MIPS 70 and 160~$\mu$m a 18.5\arcsec\/ and
40.0\arcsec aperture were used. (3): Measurements for the SMG
MM~J123633$+$6214.1. Col. (4): Instruments. Col. (5): References.}
\end{table*}

\clearpage
\begin{figure}
\begin{center}
\includegraphics[angle=270,scale=0.7]{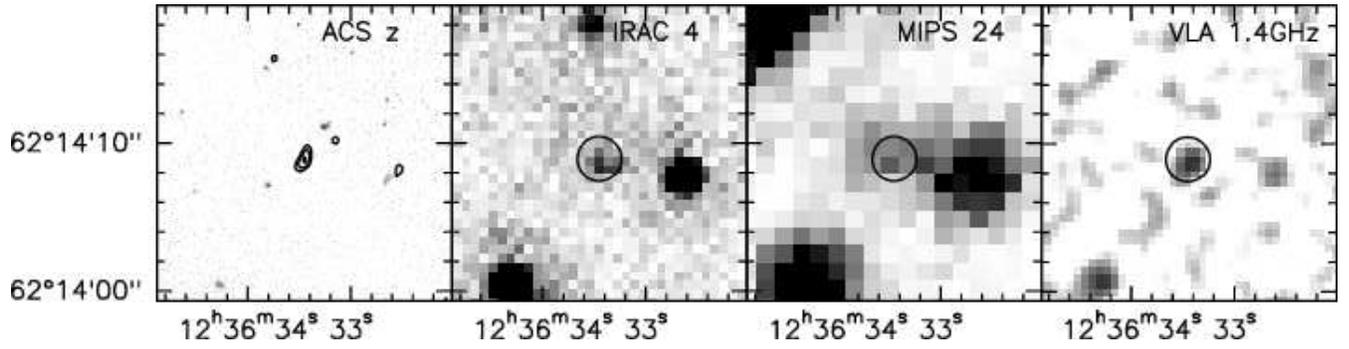}
\caption{$20\arcsec\times20\arcsec$ ACS $z_{850}$, IRAC 8.0~$\mu$m, MIPS~24~$\mu$m
  and radio images of the field of MM~J123633$+$6214.1.  PdBI contours of 1.25 continuum emission of MM~J123633$+$6214.1 overlaid on
  the  ACS $z_{850}$ image and start at 3$\sigma$ with steps of
  1~mJy. 3.0\arcsec\/ circles are drawn on the PdBI position in the IRAC 8.0~$\mu$m, MIPS~24~$\mu$m
  and radio images.}
\label{fig:cutouts}
\end{center}
\end{figure}

\end{document}